\documentclass[12 pt, amsfonts,amsmath, amssymb,color]{article}
\usepackage{authblk}
\def\baselinestretch{1.0}
\usepackage{amsmath,amssymb,amsfonts,latexsym,float,graphics,epsfig}
\usepackage{subfig}
\usepackage{verbatim}
\usepackage{relsize}
\usepackage{hyperref}
\usepackage{graphicx}
\usepackage{bm}
\usepackage{epstopdf}
\usepackage{slashed}
\usepackage{color}
\usepackage{tikz-cd}
\usepackage[utf8]{inputenc}
\usepackage[T1]{fontenc}

\def\be{\begin{equation}}
\def\ee{\end{equation}}
\def\bea{\begin{eqnarray}}
\def\eea{\end{eqnarray}}

\begin{document}

\renewcommand\theequation{\arabic{section}.\arabic{equation}}
\catcode`@=11 \@addtoreset{equation}{section}

\newtheorem{axiom}{Definition}[section]
\newtheorem{theorem}{Theorem}[section]
\newtheorem{axiom2}{Example}[section]
\newtheorem{lem}{Lemma}[section]
\newtheorem{prop}{Proposition}[section]
\newtheorem{cor}{Corollary}[section]

\newcommand{\ben}{\begin{equation*}}
\newcommand{\een}{\end{equation*}}

\let\endtitlepage\relax

\begin{titlepage}
\begin{center}
\renewcommand{\baselinestretch}{1.5}  

\vspace*{-0.5cm}

{\Large \bf{Quantum phase transitions and entanglement}}\\
 {\Large \bf{entropy in a non-Hermitian Jaynes-Cummings model}}

\vspace{9mm}
\renewcommand{\baselinestretch}{1}  

\centerline{\large{\bf Gargi Das}$^\dagger$, \large{\bf Aritra Ghosh}$^\ddagger$\footnote{{\fontsize{12pt}{14pt}\selectfont \textbf{Present Address:} School of Physics and Astronomy, Rochester Institute of Technology, Rochester, New York 14623, USA}}, \large{\bf Bhabani Prasad Mandal}$^\dagger$}

\vspace{6mm}
\normalsize
\text{$^\dagger$Department of Physics,} \\ 
\text{Banaras Hindu University,}\\
\text{Varanasi 221005, India}\\

\vspace{4mm}

\text{$^\ddagger$School of Basic Sciences,} \\ 
\text{Indian Institute of Technology Bhubaneswar,}\\
\text{Odisha 752050, India}\\

\vspace{4mm}

\text{\textbf{Email:} gargi.das@bhu.ac.in;~aritraghosh500@gmail.com;~bhabani@bhu.ac.in}

\vspace{0.4cm}

\begin{abstract}
In this paper, we describe some interesting properties of a non-Hermitian Jaynes-Cummings model. For this particular model, it is known that the Hilbert space can be described by infinitely-many two-dimensional invariant (closed) subspaces, together with the global ground state. We expose the appearance of exceptional points on such two-dimensional subspaces, together with quantum phase transitions marking the transition from real to complex eigenvalues. We also compute the spin-oscillator entanglement entropy on each invariant subspace to show that the two phases can be distinguished by their distinct entanglement-entropy profiles.
 \end{abstract}
\end{center}
\vspace*{0cm}

\end{titlepage}
\vspace*{0cm}

\clearpage

\section{Introduction}
For nearly three decades, there has been a great interest in quantum systems dictated by non-Hermitian Hamiltonians that admit real spectra. Such studies have become highly popular since the late 1990s after it was discovered that systems respecting symmetry under the joint action of parity ($\mathcal{P}$) and time reversal ($\mathcal{T}$) may admit a real spectrum \cite{Bender_1998,Bender_1999} despite the Hamiltonian being non-Hermitian. It is now understood that a consistent quantum theory with an all-real spectrum, unitary time evolution, and a probabilistic interpretation for $\mathcal{PT}$-symmetric non-Hermitian systems can be developed in a modified Hilbert space equipped with a positive-definite `$\mathcal{CPT}$' inner product \cite{Bender_2002,Bender_2007,Mostafazadeh_2010}, where $\mathcal{C}$ is an additional symmetry associated with $\mathcal{PT}$-symmetric systems. However, such a description shall fail should the $\mathcal{PT}$-symmetry be broken, in which case complex-valued eigenvalues may appear in the spectrum. On the parameter space of the system, the breaking of $\mathcal{PT}$-symmetry \cite{Khare_2000,Khare_2009,Mandal_2013,Ghatak_2013_a,Mandal_2015,Raval_2019,Znojil_2020,Mandal_2021,Modak_2021,Pal_2025} may be regarded as a quantum phase transition -- the two phases, namely, the $\mathcal{PT}$-symmetric phase and the $\mathcal{PT}$-broken phase, are separated by the so-called exceptional points \cite{Heiss_2012}. It is worth emphasizing that $\mathcal{PT}$-symmetric non-Hermitian systems have found numerous applications in various branches of physics and interdisciplinary areas \cite{Hasan_2018,Hasan_2020_b,Basu-Mallick_2001,Basu-Mallick_2004,Bender_2013,Mandal_2012,Ghatak_2015,Hajong_2024,Dwivedi_2021,Hasan_2020_a,Banerjee_2024,Hasan_2020,Hajong_2025,Brihaye_2007,Mandal_2010,Kumari_2016,Ruzicka_2021}.

\vspace{2mm}

Eventually, it has been realized that $\mathcal{PT}$-symmetric non-Hermitian systems are only a subset of a larger class of non-Hermitian systems known as pseudo-Hermitian systems which satisfy \cite{Mostafazadeh_2010,Mostafazadeh_2002_a,Mostafazadeh_2002}
  \begin{equation}\label{pseudocondition}
  H =G^{-1}H^\dagger G,\quad \quad G^\dagger=G. 
  \end{equation}
That is, the Hamiltonian and its Hermitian conjugate are related by a similarity transformation; if $G$ is the identity operator, then $H = H^\dagger$, i.e., the Hamiltonian equals its Hermitian conjugate. A Hamiltonian admitting the condition (\ref{pseudocondition}) may or may not be $\mathcal{PT}$-symmetric. It turns out that such systems may admit a real spectrum when the operator $G$ is positive-definite. In order to have a consistent quantum-mechanical picture, one must modify the Hilbert space by introducing the $G$-inner product:
\begin{equation}
(\psi, \phi)_G = \langle \psi | G | \phi \rangle. 
\end{equation}
Given the time-evolution operator $e^{-iHt}$ (we will adopt units in which $\hbar = 1$), one can write
\begin{eqnarray}
\langle \psi(t) | G | \phi(t) \rangle &=& \langle  \psi(0) | e^{iH^\dagger t} G e^{-iHt} | \phi(0) \rangle \\
&=& \langle \psi(0) | G (G^{-1} e^{iH^\dagger t} G) e^{-iHt} | \phi(0) \rangle  \nonumber \\
&=& \langle \psi(0) | G | \phi(0) \rangle,
\end{eqnarray} where in the last step, we have used the condition (\ref{pseudocondition}). The above result shows that the time evolution is unitary with respect to the modified inner product that possesses the metric $G$; because we have imposed $G$ to be positive-definite, the norms of the states are positive, allowing for a consistent quantum-mechanical interpretation. The reader is referred to \cite{Mostafazadeh_2002,Kretschmer_2004,Ghatak_2013} for more details on pseudo-Hermitian systems.  

\vspace{2mm}

As with $\mathcal{PT}$-symmetric systems that exhibit the breaking of $\mathcal{PT}$-symmetry leading to quantum phase transitions, one can also encounter similar quantum phase transitions in pseudo-Hermitian (but not necessarily $\mathcal{PT}$-symmetric) systems. In the parameter space, this happens at the exceptional points which are singular points at which two or more eigenstates (eigenvalues and eigenvectors) coalesce. Referring to the condition (\ref{pseudocondition}), the metric operator $G$ exhibits noninvertibility at an exceptional point, leading to the notion of self-orthogonality (vanishing of norm). 

\vspace{2mm}

The aim of this work is to study such quantum phase transitions in a non-Hermitian Jaynes-Cummings model. Its Hilbert space was shown earlier \cite{Mandal_2005} to be composed of an infinite number of two-dimensional invariant subspaces that are `closed', together with the global ground state, a singlet. It will be shown that on each invariant subspace, the quantum phase transition between the unbroken and broken phases can be understood as a transition between the non-dissipative and dissipative regimes. It may be observed that although the study of pseudo-Hermitian (especially, $\mathcal{PT}$-symmetric) systems has gained considerable popularity over the past decades, there have been relatively-few developments devoted to the study of information-theoretic aspects such as density matrices and entropies \cite{Fring_2019,Sinha_2024,Bagarello_2025,Ju_2019,Tzeng_2021} (see also, the general frameworks presented in \cite{Ryu_2006,Wehrl_1978,Renyi_1961}). A part of the present work is devoted to studying such a measure, namely, the spin-oscillator entanglement entropy, in order to reveal a distinction between the unbroken and broken phases in terms of their entanglement-entropy profiles.

\section{Model}\label{sec2}
\begin{figure}[h]
    \centering
    \includegraphics[height=7cm]{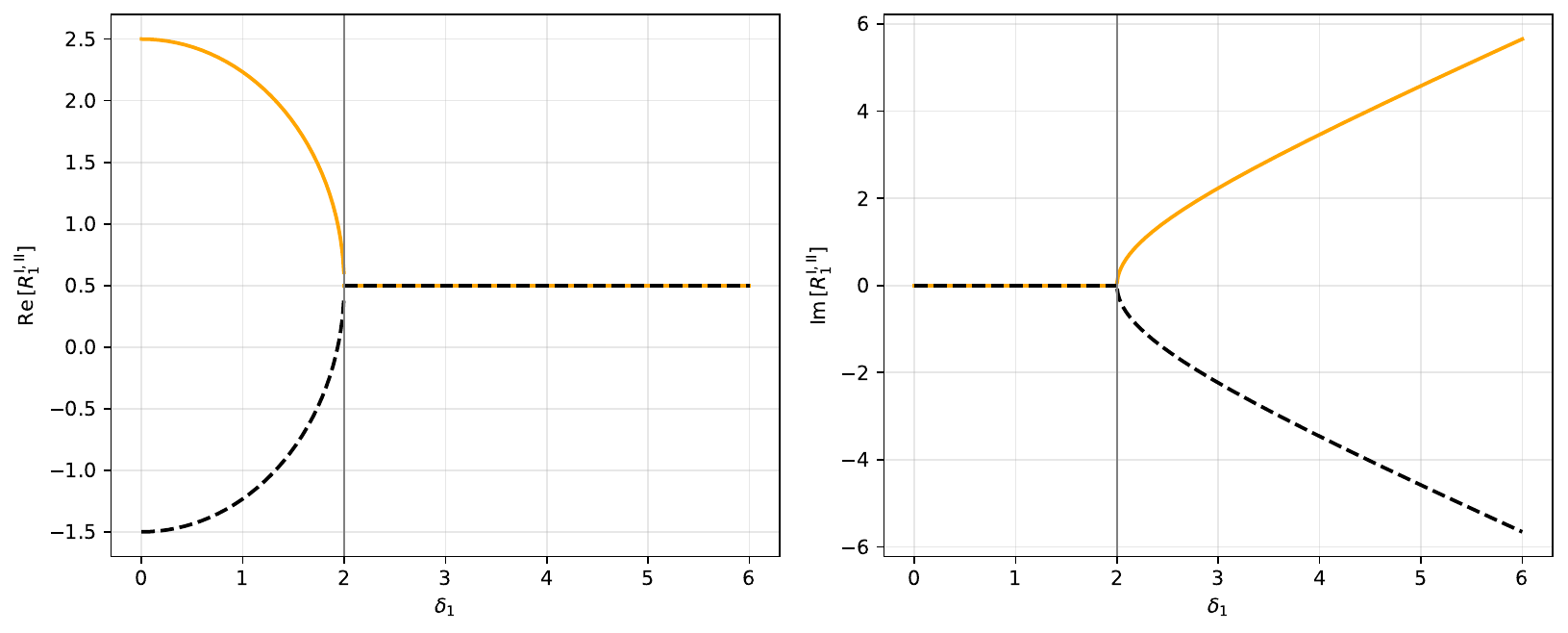}
    \caption{Eigenvalues of $H_1$ as a function of $\delta_1 = \gamma$. The eigenvalue $R^{\rm I}_1$ is indicated by the solid-orange lines while the eigenvalue $R^{\rm II}_1$ is indicated by the black-dashed lines. The vertical gray line at $\delta_1 = 2$ indicates the exceptional point.}
    \label{spectrumplot0}
\end{figure}

The system of interest is a spin-1/2 particle in an external magnetic field $\mathbf{B}$ and coupled to a bosonic oscillator through a non-Hermitian interaction \cite{Mandal_2005}. The Hamiltonian is
\begin{equation}\label{hamiltonian}
H= \mu \boldsymbol{\sigma}.{\bf B} + \omega a^\dagger a + \gamma(\sigma_{+}a - \sigma_{-}a^\dagger), 
\end{equation}
where $\boldsymbol{\sigma}=(\sigma_x,\sigma_y,\sigma_z)$ are the Pauli matrices, $\mu$ and $\gamma$ are real parameters which we shall take to be positive (the case with negative $\gamma$ can be worked out in the same way as for positive $\gamma$), $\sigma_{\pm}=\frac{1}{2}[\sigma_{x}\pm i\sigma_{y}]$ are the spin-projection operators, while $(a, a^\dagger)$ are the usual lowering and raising operators acting as 
\begin{equation}
a|n\rangle=\sqrt{n}|n-1\rangle, \quad\quad
 a^\dagger|n\rangle=\sqrt{n+1}|n+1\rangle,
\end{equation}
where the notation $|n\rangle; n = 0,1,\cdots$, for the energy eigenstates of the oscillator has been adopted. For the sake of simplicity, we will choose the external magnetic field to be in the $z$-direction, i.e., ${\bf B}=B_{0}\hat{z}$ and the Hamiltonian presented in Eq.~(\ref{hamiltonian}) is then reduced to
\begin{equation}\label{H_red}
H=\frac{\epsilon}{2}\sigma_{z} + \omega a^\dagger a + \gamma(\sigma_{+}a - \sigma_{-}a^\dagger) ,
\end{equation}
where $\epsilon = 2 \mu B_{0} $. Note that this system can also be thought of as a two-level system coupled to an oscillator, where $\epsilon$ is the spacing between the two levels. Moreover, notice that this Hamiltonian is non-Hermitian, i.e., $H \neq H^\dagger$. Nevertheless, the Hamiltonian is pseudo-Hermitian as shown earlier in \cite{Mandal_2005}. The system can be exactly solved using the standard way adopted for the Jaynes-Cummings model (see \cite{Mandal_2005} for details), i.e., the purely-number-conserving interactions ensure that the total Hilbert space splits into invariant subspaces spanned by $\left|n,\frac{1}{2}\right\rangle$ and $\left|n+1,-\frac{1}{2}\right\rangle$ in which the matrix representation of the Hamiltonian is given by (in the block basis $\{\left|n,\frac{1}{2}\right\rangle, \left|n+1,-\frac{1}{2}\right\rangle \}$)
\begin{equation}\label{H}
H_{n+1}=
\begin{pmatrix}
\frac{\epsilon}{2}+n\omega & \gamma\sqrt{n+1} \\
-\gamma\sqrt{n+1} & -\frac{\epsilon}{2}+(n+1)\omega
\end{pmatrix}.
\end{equation}
The corresponding eigenvalues are
\begin{equation}\label{energy_eigenvalues}
R^{\rm I,II}_{n+1}= \frac{1}{2}[(2n+1)\omega\pm\sqrt{(\omega-\epsilon)^2-4\gamma^2(n+1)}].
\end{equation}
Here and throughout, the label ${\rm I}$ denotes the `$+$' branch of the square-root discriminant, while ${\rm II}$ denotes the `$-$' branch. Now let us have a look at the Hermitian conjugate of $H_{n+1}$, given by
\begin{equation}\label{Hd}
H_{n+1}^\dagger=
\begin{pmatrix}
\frac{\epsilon}{2}+n\omega & -\gamma\sqrt{n+1} \\
\gamma\sqrt{n+1} & -\frac{\epsilon}{2}+(n+1)\omega
\end{pmatrix}.
\end{equation}
We will denote the eigenvalues of $H_{n+1}^\dagger$ as $L_{n+1}^{\rm I}$ and $L_{n+1}^{\rm II}$. It can then be easily seen that for the $(n+1)^{\rm th}$ invariant subspace, one gets
\begin{equation}\label{LREigen}
R^{\rm I, II}_{n+1} = L^{\rm I,II}_{n+1} =  \frac{1}{2}[(2n+1)\omega\pm\sqrt{(\omega-\epsilon)^2-4\gamma^2(n+1)}].
\end{equation}
The region of the parameter space in which the eigenvalues are real and distinct shall be called the unbroken phase. For $(\omega-\epsilon)^2-4\gamma^2(n+1) < 0$, the eigenvalues of each $H_{n+1}$ appear in complex-conjugate pairs, and this will be called the broken phase. The two above-mentioned parameter regimes are connected by the curve $(\omega-\epsilon)^2-4\gamma^2(n+1) = 0$ in the $\gamma\epsilon$-parameter space for a given $n$. For parameter values on this curve, the matrix $H_{n+1}$ becomes defective and admits only one linearly-independent eigenvector, i.e., one gets exceptional points. To summarize, 
\begin{subequations}
\begin{align}
&|\omega-\epsilon|> 2\gamma\sqrt{n+1}~~ \text{(unbroken phase)},\label{condition:unbroken}\\
&|\omega-\epsilon|< 2\gamma\sqrt{n+1}~~ \text{(broken phase)},\label{condition:broken}
\end{align}
\end{subequations}
while an exceptional point is obtained from the condition
\begin{equation}\label{critical}
\gamma_{n+1}^c= \frac{|\omega-\epsilon|}{2\sqrt{n+1}},
\end{equation}
of course, restricting to the $(n+1)^{\rm th}$ invariant subspace. In the above, the superscript `c' on $\gamma$ indicates its value at which, for given $n$, $\omega$, and $\epsilon$, one hits an exceptional point. The eigenvalues $R^{\rm I,II}_{n+1}$ are depicted in Fig. (\ref{spectrumplot0}) for $n=0$. 

\vspace{2mm}

\begin{figure}[htbp]
    \centering
    \subfloat[$n=0$\label{fig:11a}]{
        \includegraphics[width=0.45\linewidth]{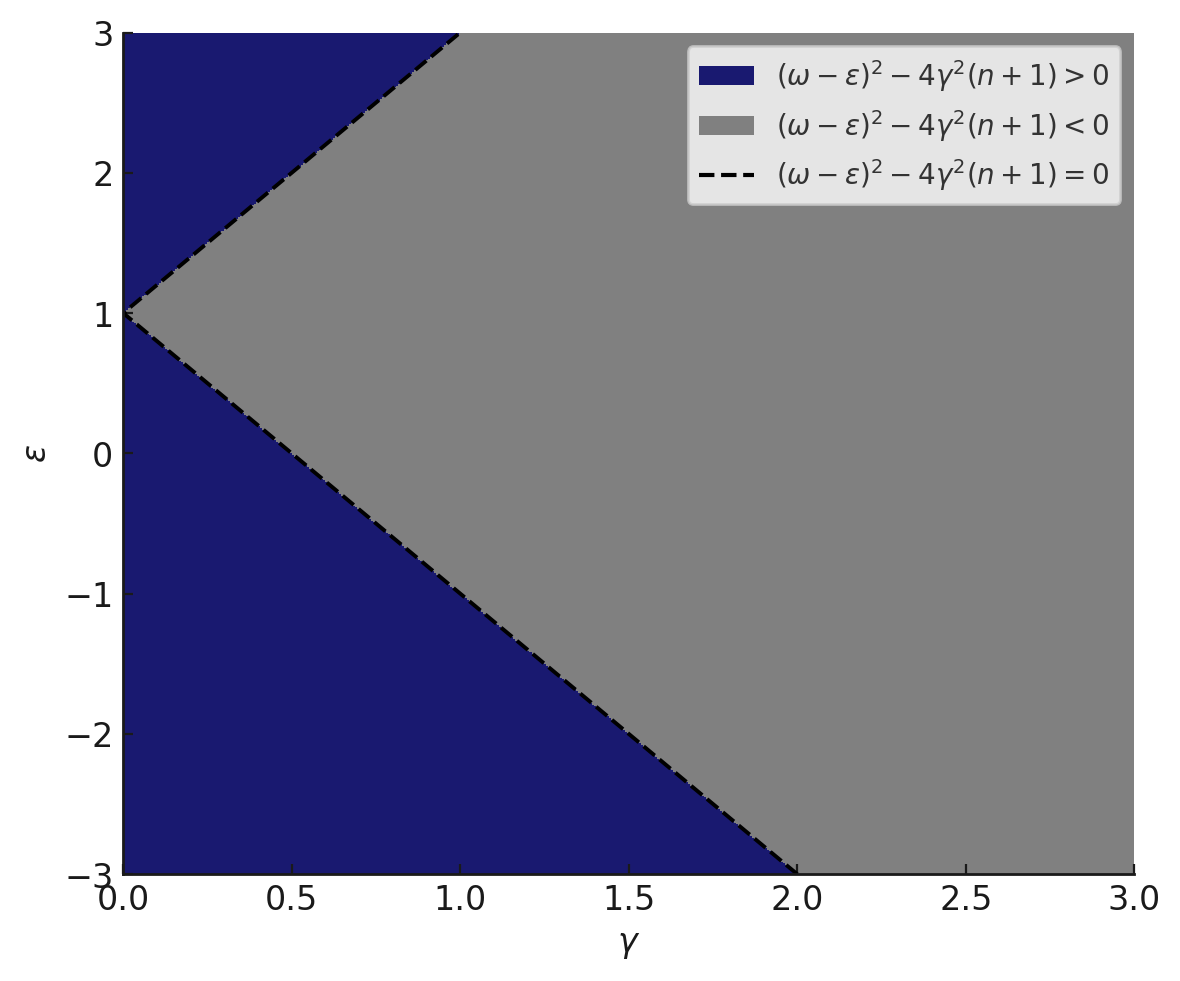}
    }\hfill
    \subfloat[$n=1$\label{fig:11b}]{
        \includegraphics[width=0.45\linewidth]{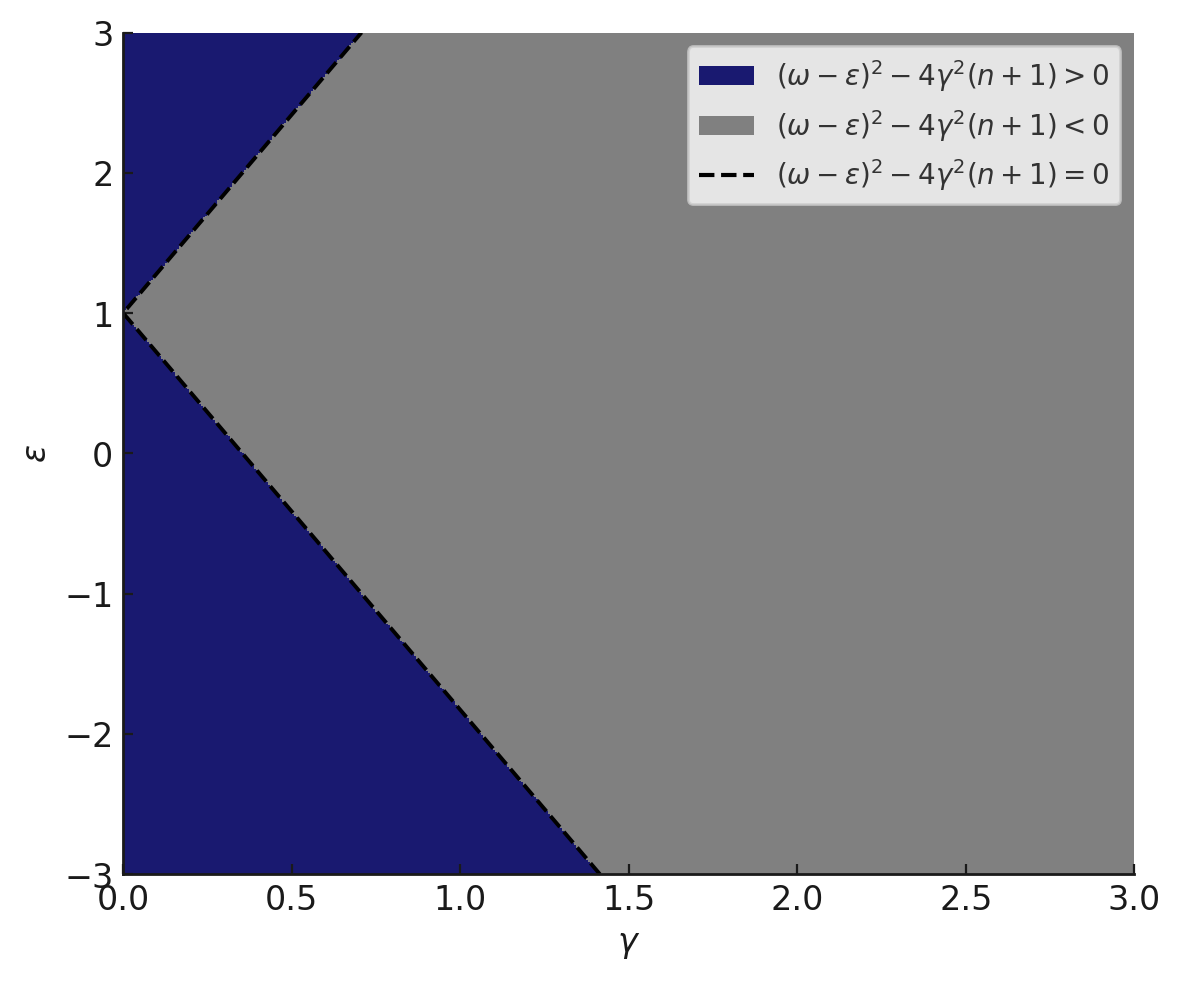}
    }

    \vspace{0.2cm}

    \subfloat[$n=2$\label{fig:11c}]{
        \includegraphics[width=0.45\linewidth]{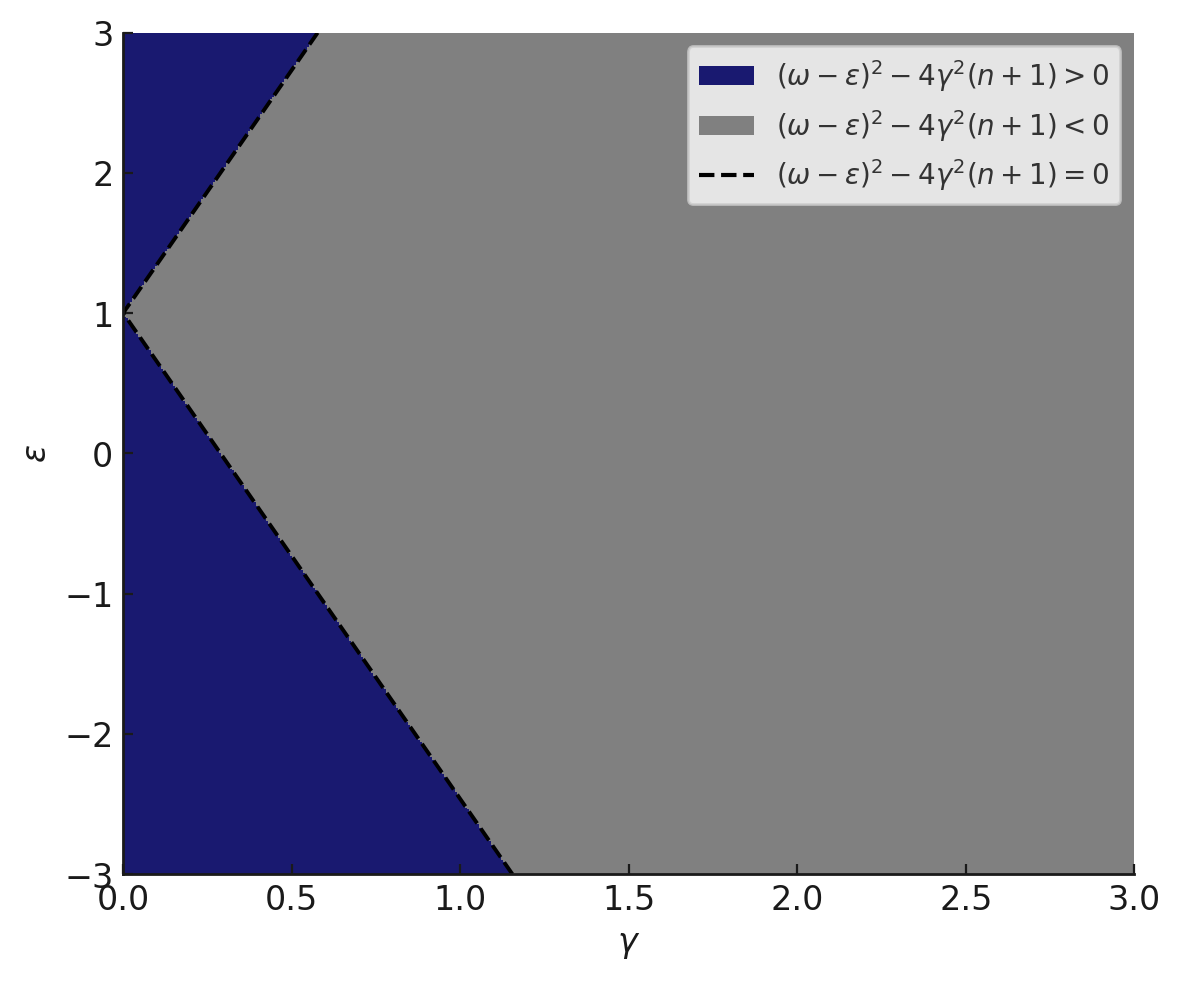}
    }\hfill
    \subfloat[$n=3$\label{fig:11d}]{
        \includegraphics[width=0.45\linewidth]{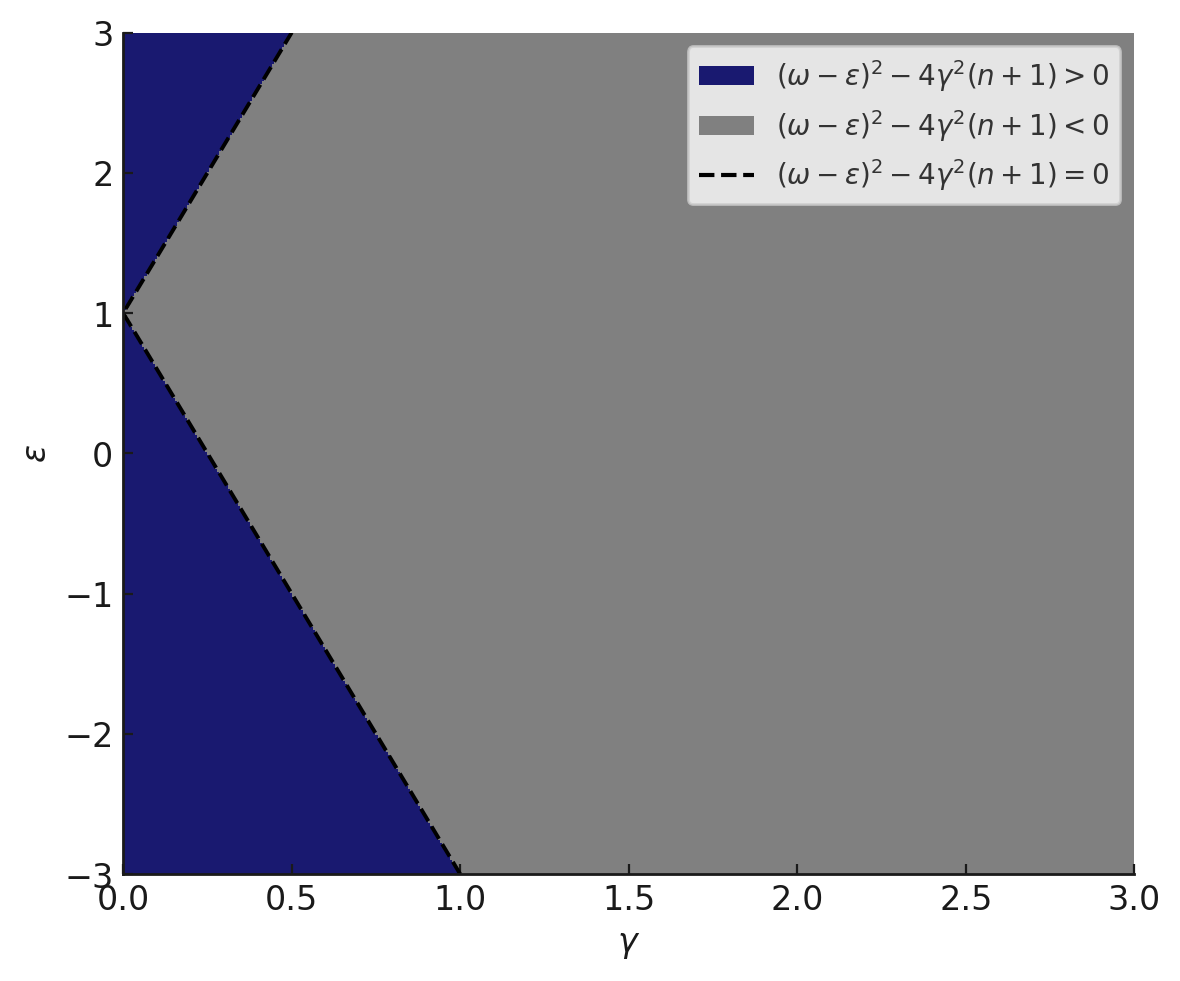}
    }

    \caption{
        Figures showing the broken (gray) and unbroken (blue) phases in the $\gamma\epsilon$-parameter space for $\omega = 1$ and with $n=0,1,2,3$. Exceptional points lie on the dashed lines separating the two regions.
    }
    \label{fig:11}
\end{figure}

As noted earlier, the transition between the unbroken and broken phases through the exceptional point can be understood as a quantum phase transition. In Fig. (\ref{fig:11}), we have plotted the equation $(\omega-\epsilon)^2-4\gamma^2(n+1) = 0$, obtained by setting to zero the discriminant appearing in the roots shown in Eq. (\ref{LREigen}) for $n=0,1,2,3$. The regions shaded in gray correspond to negative values of the discriminant giving rise to eigenvalues of $H_{n+1}$ that are complex conjugates. On the other hand, the regions shaded in blue correspond to positive values of the discriminant giving rise to real and distinct eigenvalues of $H_{n+1}$. These two regions, therefore, correspond to the broken and unbroken phases, respectively, with the lines separating them (discriminant = 0) corresponding to the exceptional points in the $\gamma\epsilon$-parameter space. 

\section{The metric and intertwined Hamiltonians}\label{sec3}
For a non-Hermitian Hamiltonian $H$, there are distinct left and right eigenvectors $\left\langle L_m \right|$ and $\left| R_m \right\rangle$, respectively. In the biorthogonal interpretation of quantum mechanics \cite{Brody_2014}, the inner product is defined as $\left\langle L_{m_1} | R_{m_2} \right\rangle = \delta_{m_1,m_2} $. A metric $G$ can always be chosen as (except at an exceptional point) \cite{Hajong_2024,Shukla_2023}
\begin{equation}\label{Ggeneraldef}
G = \sum_m \left| L_m \right\rangle \left\langle L_m \right|,
\end{equation}
where $\left| L_m \right\rangle$ are the eigenstates of $H^\dagger$, i.e.,
\begin{equation}\label{vn}
H | R_m \rangle =  R_m | R_m \rangle , \quad H^\dagger | L_m \rangle =  L_m | L_m \rangle .
\end{equation}
The inner product may now be defined as
\begin{equation}
\left\langle R_{m_1} \right| \mathcal{O} \left| R_{m_2} \right\rangle_G = \left\langle R_{m_1} \right| G \mathcal{O} \left| R_{m_2} \right\rangle,
\end{equation} where $\mathcal{O}$ is some observable. Substituting Eq. (\ref{Ggeneraldef}), one immediately obtains
\begin{equation}
\left\langle R_{m_1} \right| \mathcal{O} \left| R_{m_2} \right\rangle_G = \left\langle L_{m_1} \right| \mathcal{O} \left| R_{m_2} \right\rangle,
\end{equation} where the biorthogonal inner product appears naturally. Notice that we have not made use of the condition (\ref{pseudocondition}), i.e., the above-mentioned prescription for constructing the metric operator can also be used in the broken phase. In that phase, however, the operator constructed from Eq. (\ref{Ggeneraldef}) should be understood as a positive biorthogonal metric defining the inner product, and not in the sense of the condition (\ref{pseudocondition}). In this sense, the notions of metrics appearing in Eqs. (\ref{pseudocondition}) and (\ref{Ggeneraldef}) are generally distinct -- while the former ensures that the non-Hermitian Hamiltonian is self-adjoint in a modified inner product, the latter's role is just to ensure the positivity of the norm. In particular, the two notions cannot agree in the broken phase, for that would imply a unitary evolution in the phase with complex eigenvalues. In the ensuing analysis, we shall refer to the latter sense of the metric defined in Eq. (\ref{Ggeneraldef}). 

\vspace{2mm}

While the metric operator $G$ that allows one to define a consistent inner product can be defined both in the unbroken and broken phases, these two phases are distinct due to the distinct nature of their spectra -- in the unbroken phase, the eigenvalues are real while in the broken phase, the eigenvalues appear in complex-conjugate pairs, as noticed earlier. Since the same underlying Hamiltonian leads to these two distinct behaviors in different regions of the parameter space, it is natural to inquire whether in each of these phases, the Hamiltonian can be mapped via intertwiners to new representations that can give rise to the real/complex-conjugate eigenvalues. This is true for the unbroken phase where the operator $g = \sqrt{G}$ acts as the intertwiner, leading to \cite{Mostafazadeh_2010}
\begin{equation}
h = g H g^{-1},
\end{equation} which is Hermitian and isospectral to $H$. Note that while $GH$ is Hermitian by virtue of Eq. (\ref{pseudocondition}), it is not isospectral to $H$. In the broken phase, denoting the corresponding quantities with a `tilde', $\tilde{G}$ is still positive-definite by construction and thus $\tilde{g} = \sqrt{\tilde{G}}$ can act as an intertwiner and the matrix $\tilde{h} = \tilde{g} H \tilde{g}^{-1}$, while isospectral to $H$, is not Hermitian. With the understanding that the metric operator can be constructed using the general prescription given in Eq. (\ref{Ggeneraldef}), we will now proceed to study the two phases mentioned earlier.

\subsection{Unbroken phase}
In this phase, the condition given in Eq. (\ref{condition:unbroken}) is met and the eigenvalues $R^{\rm I,II} = L^{\rm I, II}$ are real and distinct. Let us find the normalized eigenvectors of the matrix $H_{n+1}$. For definiteness, let us set the parameters $\omega=1$ and $\epsilon=5$ in all the following discussions. Keeping in mind that a non-Hermitian matrix admits distinct left and right eigenvectors, we will use the notations
\begin{eqnarray}\label{eigenvector}
\text{Right eigenvectors:}\quad
|R^{\rm I}_{n+1}\rangle &=& -\frac{1}{\sqrt{ 1-(\Phi^{\rm I})^{2} }}
\begin{pmatrix} 1\\[2pt] \Phi^{\rm I}\end{pmatrix}, \\
|R^{\rm II}_{n+1}\rangle &=& -\frac{1}{\sqrt{ (\Phi^{\rm II})^{2}-1 }}
\begin{pmatrix} 1\\[2pt] \Phi^{\rm II}\end{pmatrix},\\
\text{Left eigenvectors:}\quad
|L^{\rm I}_{n+1}\rangle  &=& \frac{1}{\sqrt{ 1-(\Phi^{\rm I})^{2} }}
\begin{pmatrix} -1\\[2pt] \Phi^{\rm I}\end{pmatrix},\\
|L^{\rm II}_{n+1}\rangle &=& \frac{1}{\sqrt{ (\Phi^{\rm II})^{2}-1 }}
\begin{pmatrix} 1\\[2pt] -\Phi^{\rm II}\end{pmatrix},
\end{eqnarray}
where $\Phi^{\rm I,II} = \frac{\pm \sqrt{4-\delta_{n+1}^2} - 2}{\delta_{n+1}}$ and $\delta_{n+1} = \sqrt{(n+1)}\gamma$. In the $(n+1)^{\rm th}$ invariant subspace, one can calculate the metric in order to define the inner product as described earlier. The metric turns out to be
\begin{equation}\label{G}
G_{n+1} = \sum_{i={\rm I, II}} |L^i_{n+1}\rangle\langle L^i_{n+1}|= \frac{1}{\sqrt{4-\delta_{n+1}^2}}
\begin{pmatrix}
2 & \delta_{n+1}  \\
\delta_{n+1}  & 2 
\end{pmatrix},
\end{equation}
where $\delta_{n+1}^2 < 4$ represents the unbroken phase. Thus the intertwiner reads
\begin{equation}\label{g_n}
g_{n+1}=\sqrt{G_{n+1}}
=\frac{1}{2(4-\delta_{n+1}^{2})^{1/4}}
\begin{pmatrix}
\sqrt{2+\delta_{n+1}}+\sqrt{2-\delta_{n+1}} & \sqrt{2+\delta_{n+1}}-\sqrt{2-\delta_{n+1}}\\[6pt]
\sqrt{2+\delta_{n+1}}-\sqrt{2-\delta_{n+1}} & \sqrt{2+\delta_{n+1}}+\sqrt{2-\delta_{n+1}}
\end{pmatrix},
\end{equation}
while its inverse is
\begin{equation}
g_{n+1}^{-1}
=\frac{(4-\delta_{n+1}^{2})^{1/4}}{2}
\begin{pmatrix}
\frac{1}{\sqrt{2+\delta_{n+1}}}+\frac{1}{\sqrt{2-\delta_{n+1}}} &
\frac{1}{\sqrt{2+\delta_{n+1}}}-\frac{1}{\sqrt{2-\delta_{n+1}}} \\[10pt]
\frac{1}{\sqrt{2+\delta_{n+1}}}-\frac{1}{\sqrt{2-\delta_{n+1}}} &
\frac{1}{\sqrt{2+\delta_{n+1}}}+\frac{1}{\sqrt{2-\delta_{n+1}}}
\end{pmatrix}.
\end{equation}
Using these, one can now find the Hermitian equivalent of the Hamiltonian $H_{n+1}$, of course, restricted to $\delta_{n+1}^2 = (n+1)\gamma^2 < 4$ for the present choice of the parameters. One gets 
\begin{equation}\label{h_n}
h_{n+1} = g_{n+1} H_{n+1} g_{n+1}^{-1}  = \begin{pmatrix}
n+\frac{1}{2}+\sqrt{ 4-(n+1)\gamma^{2} } & 0 \\[6pt]
0 & n+\frac{1}{2}-\sqrt{ 4-(n+1)\gamma^{2} }
\end{pmatrix},
\end{equation}
which is diagonal and admits the same eigenvalues as $H_{n+1}$ for $\omega = 1$, $\epsilon = 5$, and $\delta_{n+1}^2 = (n+1)\gamma^2 < 4$, the latter being chosen as $\delta_{n+1} = +\sqrt{n+1}\gamma$ in the above-mentioned calculations.

\subsection{Broken phase}
In the broken phase, using Eq. (\ref{condition:broken}), the eigenvectors are
\begin{eqnarray}\label{eigenvectord}
\text{Right eigenvectors:}\quad
|\tilde R^{\rm I}_{n+1}\rangle &=& \frac{1}{\sqrt{ (\tilde{\Phi}^{\rm I})^{2}-1 }}
\begin{pmatrix} 1\\[2pt] \tilde{\Phi}^{\rm I}\end{pmatrix},\\
|\tilde R^{\rm II}_{n+1}\rangle &=& \frac{1}{\sqrt{ (\tilde{\Phi}^{\rm II})^{2}-1 }}
\begin{pmatrix} 1\\[2pt] \tilde{\Phi}^{\rm II}\end{pmatrix},\\
\text{Left eigenvectors:}\quad
|\tilde{L}^{\rm I}_{n+1}\rangle  &=& \frac{1}{\sqrt{ (\tilde{\Phi}^{\rm II})^{2}-1 }}
\begin{pmatrix} -1\\[2pt] \tilde{\Phi}^{\rm II}\end{pmatrix},\\
|\tilde{L}^{\rm II}_{n+1}\rangle &=& \frac{1}{\sqrt{ (\tilde{\Phi}^{\rm I})^{2}-1 }}
\begin{pmatrix} -1\\[2pt] \tilde{\Phi}^{\rm I}\end{pmatrix},
\end{eqnarray}
where $\tilde{\Phi}^{\rm I,II} = \frac{-2 \pm i\sqrt{\delta_{n+1}^2-4}}{\delta_{n+1}}$. We have adopted the convention that a symbol with `tilde' represents the corresponding quantity in the broken phase. The metric is calculated from Eq. (\ref{Ggeneraldef}) to be
\begin{equation}\label{Gd}
\tilde{G}_{n+1} = \sum_{i = {\rm I, II}} |\tilde{L}^i_{n+1}\rangle\langle \tilde{L}^i_{n+1}| = 
\frac{1}{\sqrt{\delta_{n+1}^2-4}}
\begin{pmatrix}
\delta_{n+1} & 2  \\
2  & \delta_{n+1} 
\end{pmatrix},
\end{equation}
which is Hermitian and positive-definite for $\delta_{n+1}^2 > 4$. The intertwiner is found to be
\begin{equation}\label{tildeg_n}
\tilde{g}_{n+1}=\sqrt{\tilde{G}_{n+1}}
=\frac{1}{2(\delta_{n+1}^{2}-4)^{1/4}}
\begin{pmatrix}
\sqrt{\delta_{n+1}+2}+\sqrt{\delta_{n+1}-2} & \sqrt{\delta_{n+1}+2}-\sqrt{\delta_{n+1}-2}\\[6pt]
\sqrt{\delta_{n+1}+2}-\sqrt{\delta_{n+1}-2} & \sqrt{\delta_{n+1}+2}+\sqrt{\delta_{n+1}-2}
\end{pmatrix},
\end{equation} and its inverse is
\begin{equation}
\tilde{g}_{n+1}^{-1}
=\frac{(\delta_{n+1}^{2}-4)^{1/4}}{2}
\begin{pmatrix}
\frac{1}{\sqrt{\delta_{n+1}+2}}+\frac{1}{\sqrt{\delta_{n+1}-2}} &
\frac{1}{\sqrt{\delta_{n+1}+2}}-\frac{1}{\sqrt{\delta_{n+1}-2}} \\[10pt]
\frac{1}{\sqrt{\delta_{n+1}+2}}-\frac{1}{\sqrt{\delta_{n+1}-2}} &
\frac{1}{\sqrt{\delta_{n+1}+2}}+\frac{1}{\sqrt{\delta_{n+1}-2}}
\end{pmatrix}.
\end{equation}
Thus the isospectral Hamiltonian turns out to be
\begin{equation}\label{tildeh_n}
\tilde{h}_{n+1} = \tilde{g}_{n+1} H_{n+1} \tilde{g}_{n+1}^{-1} = \begin{pmatrix}
n+\frac{1}{2} & \sqrt{(n+1)\gamma^{2}-4}\\[6pt]
-\sqrt{(n+1)\gamma^{2}-4} & n+\frac{1}{2}
\end{pmatrix},
\end{equation}
where $\omega = 1$, $\epsilon = 5$, and $\delta_{n+1}^2 = (n+1)\gamma^2 > 4$, the latter being chosen as $\delta_{n+1} = +\sqrt{n+1}\gamma$. This Hamiltonian is non-Hermitian and admits the same eigenvalues as $H_{n+1}$ in the broken phase for the above-mentioned parameter values.

\section{Diagnostics of the phase transition}\label{sec4}
\subsection{Signatures of the phase transition}
Let us now assess the signatures of the aforementioned quantum phase transition on each invariant subspace. Considering the $(n+1)^{\rm th}$ invariant subspace, let us introduce the normalized biorthogonal pure-state projectors
\begin{equation}
\rho^{\rm I,II}_{n+1} = \frac{|R^{\rm I,II}_{n+1}\rangle\langle L^{\rm I,II}_{n+1}|}{\langle L^{\rm I,II}_{n+1}|R^{\rm I,II}_{n+1}\rangle},
\label{projector_def}
\end{equation}
for the unbroken phase and with tildes on the corresponding quantities for the broken phase. For the unbroken phase, a direct calculation starting with the matrix $H_{n+1}$ given in Eq. (\ref{H}) and taking $\omega = 1$, $\epsilon = 5$, and $\delta_{n+1}^2 = (n+1)\gamma^2$ gives the expressions 
\begin{equation}
\rho^{\rm I}_{n+1} = \frac{1}{2\sqrt{4 - \delta_{n+1}^2}}
\begin{pmatrix}
\sqrt{4 - \delta_{n+1}^2} + 2 & \delta_{n+1} \\
-\delta_{n+1} & \sqrt{4 - \delta_{n+1}^2} - 2
\end{pmatrix}, 
\end{equation}
\begin{equation}
\rho^{\rm II}_{n+1}= \frac{1}{2\sqrt{4 - \delta_{n+1}^2}}
\begin{pmatrix}
\sqrt{4 - \delta_{n+1}^2} - 2 & -\delta_{n+1} \\
\delta_{n+1} & \sqrt{4 - \delta_{n+1}^2} + 2
\end{pmatrix}.
\end{equation}
These projectors are non-Hermitian but satisfy ${\rm Tr} [\rho^{\rm I,II}_{n+1}] = 1$, as well as $(\rho^{\rm I,II}_{n+1})^2 = \rho^{\rm I,II}_{n+1}$. In the broken phase, a similar calculation leads to the expressions 
\begin{equation}
\tilde{\rho}^{\rm I}_{n+1}= \frac{1}{2i\sqrt{\delta_{n+1}^2 - 4}}
\begin{pmatrix}
i\sqrt{\delta_{n+1}^2 - 4} + 2 & \delta_{n+1} \\
-\delta_{n+1} & i\sqrt{\delta_{n+1}^2 - 4} - 2
\end{pmatrix}, 
\end{equation}
\begin{equation}
\tilde{\rho}^{\rm II}_{n+1} = \frac{1}{2i\sqrt{\delta_{n+1}^2 - 4}}
\begin{pmatrix}
i\sqrt{\delta_{n+1}^2 - 4} - 2 & -\delta_{n+1} \\
\delta_{n+1} & i\sqrt{\delta_{n+1}^2 - 4} + 2
\end{pmatrix}.
\end{equation}
The projectors satisfy ${\rm Tr} [\tilde{\rho}^{\rm I,II}_{n+1}] = 1$ and $(\tilde{\rho}^{\rm I,II}_{n+1})^2 = \tilde{\rho}^{\rm I,II}_{n+1}$. One can observe at once that at the exceptional points marking the transition between the two phases, these projectors become singular, scaling near $\delta_{n+1}^2 \rightarrow (\delta_{n+1}^c)^2 = 4$ as $(|\delta_{n+1}^2 - (\delta_{n+1}^c)^2|)^{-1/2}$, i.e., as $(|\delta_{n+1} - \delta_{n+1}^c|)^{-1/2}$. A similar divergence can be found in the metric operators of the two phases, i.e., 
\begin{eqnarray}
G_{n+1} \sim \big[- (\delta_{n+1}^2 - 4)\big]^{-1/2}, \quad \quad \delta_{n+1}^2 \rightarrow 4^-, \\
\tilde{G}_{n+1} \sim \big[+(\delta_{n+1}^2 - 4)\big]^{-1/2}, \quad \quad \delta_{n+1}^2 \rightarrow 4^+. 
\end{eqnarray}
One thus encounters a critical exponent of 1/2 with which the metric operators and the biorthogonal projectors diverge at the point of the phase transition. This exponent originates from the square-root singularity at second-order exceptional points \cite{Heiss_2012}. 

\subsection{Coherence to decoherence transition}
We shall now demonstrate how the phase transition can be understood as a coherence to decoherence transition on each invariant subspace. Considering the $(n+1)^{\rm th}$ invariant subspace, the $2 \times 2$ Hamiltonian matrix is `similar' (except at $(n+1)\gamma^2 = 4$) to
\begin{eqnarray}
\mathcal{H}_{n+1} &=& \begin{pmatrix}
n+\frac{1}{2} & \sqrt{(n+1)\gamma^{2}-4}\\[4pt]
-\sqrt{(n+1)\gamma^{2}-4} & n+\frac{1}{2}
\end{pmatrix} \nonumber\\
&=& \left(n + \frac{1}{2} \right) \mathbb{I}_2 + i \Gamma \sigma_y, \label{calH_n}
\end{eqnarray} where $\mathbb{I}_2$ is the $2 \times 2$ identity matrix, $\Gamma = \sqrt{(n+1)\gamma^{2}-4}$, and we have picked $\omega =1$ and $\epsilon = 5$. This matrix is identical to $\tilde{h}_{n+1}$ found in Eq. (\ref{tildeh_n}) for the broken phase and so it is isospectral to it for $(n+1)\gamma^{2} > 4$. For the unbroken phase, the diagonal isospectral matrix $h_{n+1}$ presented in Eq. (\ref{h_n}) is similar to $\mathcal{H}_{n+1}$ quoted above by the unitary rotation
\begin{equation}
\mathcal{H}_{n+1} = U h_{n+1} U^{-1}, \quad \quad U = e^{-i\frac{\pi}{4}\sigma_x},
\end{equation} for $(n+1)\gamma^{2} < 4$ and upon defining $\Gamma = i\Lambda$ with $\Lambda = \sqrt{4 - (n+1)\gamma^{2}}$ to enforce analytic continuation of $\mathcal{H}_{n+1}$ in the unbroken phase. Thus $\mathcal{H}_{n+1}$ furnishes an isospectral representative of $H_{n+1}$ [Eq. (\ref{H}); $\omega =1$, $\epsilon = 5$] in both the phases, except at the exceptional point where $H_{n+1}$ loses diagonalizability. It should be particularly emphasized that the phase-specific intertwiners given in Eqs. (\ref{g_n}) and (\ref{tildeg_n}) constitute two analytic branches separated by the exceptional-point singularity. Hence the similarity transforms they define cannot be extended through the exceptional point without encountering divergence, the latter is precisely the origin of the critical behavior. It should be clarified that $\sigma_x$ and $\sigma_y$ used above denote the standard Pauli matrices in the effective $2\times2$ representation of the invariant subspace spanned by $\{|n,\frac{1}{2}\rangle, |n+1,-\frac{1}{2}\rangle\}$, i.e., they act on this subspace and should not be identified directly with the physical spin-Pauli operators on the full Jaynes-Cummings Hilbert space.

\vspace{2mm}

Referring to $\mathcal{H}_{n+1}$ appearing in Eq. (\ref{calH_n}), since the anti-Hermitian part is not negative-semidefinite, it cannot be obtained from a Lindblad master equation via a completely-positive and trace-preserving Liouvillian \cite{Plenio_1998}. However, the above-mentioned form can describe a two-level system and one can identify dissipative and non-dissipative regimes. Let us consider the master equation (no jump terms)
\begin{equation}
\frac{d\rho}{dt} = - i\big(\mathcal{H}_{n+1}\rho - \rho \mathcal{H}_{n+1}^\dagger\big)
\quad \implies \quad
\rho(t) = e^{-i\mathcal{H}_{n+1}t} \rho(0) e^{+i\mathcal{H}_{n+1}^\dagger t}.
\end{equation}
Since $[\mathbb{I}_2,\sigma_y]=0$, the expression for $\rho(t)$ can be simplified by cancelling the global phase factor to give (assuming $\Gamma \in \mathbb{R}$)
\begin{equation}
\rho(t) = S(t) \rho(0) S(t),
\quad \quad
S(t) =  e^{\Gamma t \sigma_y}.
\label{S_def}
\end{equation}
In order to quantify the transition between the regimes with coherent dynamics and that with decoherence, let us define the Bloch representation $\rho(0)=\frac{1}{2}\big(\mathbb{I}_2+ \mathbf r\!\cdot\!\boldsymbol{\sigma}\big)$ with Bloch vector $\mathbf{r}=(r_x,r_y,r_z)$. We thus have $S(t) = \cosh (\Gamma t) \mathbb{I}_2 + \sinh (\Gamma t) \sigma_y$ upon using the Pauli-matrix exponential identity, and with some algebraic manipulations, we can write
\begin{equation}
\rho(t)=\frac{1}{2}\Big[\big(\cosh (2\Gamma t) + r_y \sinh (2\Gamma t)\big) \mathbb{I}_2 + r_x \sigma_x + r_z \sigma_z + \big(\sinh (2\Gamma t) + r_y \cosh (2\Gamma t)\big)\sigma_y
\Big].
\end{equation}
The above can be normalized by dividing with the factor $D(t) = {\rm Tr} [\rho(t)] = \cosh(2\Gamma t) + r_y \sinh(2\Gamma t)$, which is the no-jump survival probability. We can thus obtain the following picture:
\begin{enumerate}
\item When $\Gamma \in \mathbb{R}$, i.e., in the broken phase with complex-conjugate eigenvalues, the density matrix undergoes a non-conservative evolution due to the appearance of the hyperbolic sines and cosines. This encodes decoherence in the no-jump evolution.
\item When $\Gamma = i \Lambda$ with $\Lambda \in \mathbb{R}$, i.e., in the unbroken phase, the hyperbolic sines and cosines reduce to trigonometric (circular) sines and cosines, thereby leading to oscillatory dynamics. Eq. (\ref{S_def}) is then modified to $\rho(t) = S(t) \rho(0) S^\dagger(t)$, where $S(t) = e^{i\Lambda t \sigma_y}$ is unitary and the transformation $\rho(t)=S(t)\rho(0)S^\dagger(t)$ is trace-preserving. 
\end{enumerate}
In other words, the quantum phase transition on the $(n+1)^{\rm th}$ invariant subspace between the unbroken and broken phases can be interpreted as a coherence to decoherence transition of a qubit. This behavior can be compared with the similar coherence-decoherence transition studied for dissipative quantum-oscillatory systems in contact with heat baths \cite{Egger_1997,Bandyopadhyay_2006}. 

\section{Entanglement entropy on the $(n+1)^{\rm th}$ invariant subspace}\label{sec5}
Let us now quantify the spin-oscillator entanglement by assessing the entanglement entropy between these two degrees of freedom upon restricting to the $(n+1)^{\rm th}$ invariant subspace, spanned by the states $\left|n,\frac{1}{2}\right\rangle$ and $\left|n+1,-\frac{1}{2}\right\rangle$. The Hamiltonian matrix is given by (in the block basis $\{\left|n,\frac{1}{2}\right\rangle, \left|n+1,-\frac{1}{2}\right\rangle \}$)
\begin{equation}
H_{n+1} = \begin{pmatrix}
n+\frac{5}{2} & \delta_{n+1}\\[3pt]
-\delta_{n+1} & n-\frac{3}{2}
\end{pmatrix},
\end{equation}
where we have picked $\omega=1$, $\epsilon=5$, and $\delta_{n+1}^2 = (n+1)\gamma^2$. Let the right eigenvectors be $|R^{\rm I,II}_{n+1} \rangle$ and the left ones be $\langle L^{\rm I,II}_{n+1} |$, as considered earlier. In the $\{\left|n,\frac{1}{2}\right\rangle, \left|n+1,-\frac{1}{2}\right\rangle \}$ basis, the right eigenvectors are found to be (up to normalization)
\begin{equation}
|R^{\rm I,II}_{n+1}\rangle \propto \left|n,\frac{1}{2}\right\rangle + \alpha^{\rm I,II}_{n+1} \left|n+1,-\frac{1}{2}\right\rangle,
\end{equation}
where $\alpha^{\rm I,II}_{n+1} = \frac{-2 \pm \sqrt{4 - \delta_{n+1}^2}}{\delta_{n+1}}$. Similarly, the left eigenvectors give (up to normalization and Hermitian conjugation)
\begin{equation}
|L^{\rm I,II}_{n+1}\rangle \propto \left|n,\frac{1}{2}\right\rangle - \alpha^{\rm I,II}_{n+1} \left|n+1,-\frac{1}{2}\right\rangle.
\end{equation} 
The normalization condition for the biorthogonal picture is $\langle L^i_{n+1} | R^j_{n+1} \rangle = \delta_{i, j}$, with $i,j \in \{{\rm I,II}\}$, i.e., the normalization factor is $\sqrt{1 - (\alpha^{\rm I,II}_{n+1})^2}$, which is not well defined for $(\alpha^{\rm I,II}_{n+1})^2 = 1$. One therefore has\footnote{\textbf{Post-publication clarification:} The expressions displayed here for the left eigenvectors and the biorthogonal normalization are intended only schematically. The quoted biorthogonal-normalization factor should be understood as a coefficient-level shorthand rather than as a single biorthogonal-normalization formula valid uniformly in both the broken and unbroken phases and for both eigenvector branches. In the broken phase, in particular, each left eigenvector involves the coefficient carrying the branch label opposite to its own. This refinement does not affect the entropy calculation, which relies only on Dirac-normalized right eigenvectors.}
\begin{eqnarray}
|R^{\rm I,II}_{n+1}\rangle &=& \frac{1}{\sqrt{1 - (\alpha^{\rm I,II}_{n+1})^2}} \left|n,\frac{1}{2}\right\rangle + \frac{\alpha^{\rm I,II}_{n+1}}{\sqrt{1 - (\alpha^{\rm I,II}_{n+1})^2}} \left|n+1,-\frac{1}{2}\right\rangle, \\
|L^{\rm I,II}_{n+1}\rangle &=& \frac{1}{\sqrt{1 - (\alpha^{\rm I,II}_{n+1})^2}} \left|n,\frac{1}{2}\right\rangle - \frac{\alpha^{\rm I,II}_{n+1}}{\sqrt{1 - (\alpha^{\rm I,II}_{n+1})^2}} \left|n+1,-\frac{1}{2}\right\rangle. 
\end{eqnarray}
For computing the spin-oscillator entanglement entropy, one can consider tracing over the bosonic degree of freedom on the orthogonal projectors defined in Eq. (\ref{projector_def}). However, it is noteworthy that the density matrix cannot be guaranteed to be positive-definite in the biorthogonal prescription when considering the broken phase. In order to avoid this, we shall choose to work with the so-called Dirac-normalization scheme (as opposed to the biorthogonal scheme) in which the norm of a state $|R^{\rm I,II}_{n+1} \rangle$ shall be defined to be the Dirac norm $\langle R^{\rm I,II}_{n+1} | R^{\rm I,II}_{n+1} \rangle$ which is positive-definite \cite{Ruzicka_2021,Sinha_2024}. In fact, the use of the Dirac-normalization scheme for studying entropic measures for non-Hermitian systems has been suggested earlier (see for example, \cite{Sinha_2024}). 

\vspace{2mm}

\begin{figure}[h]
    \centering
    \includegraphics[height=9.7cm]{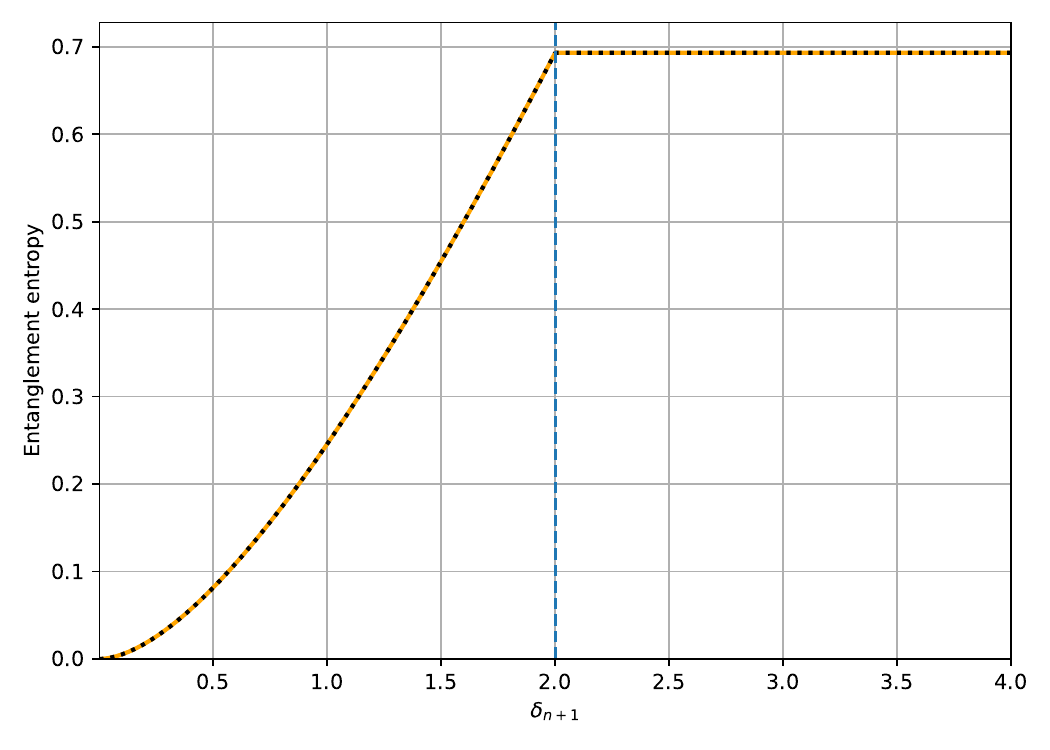}
    \caption{Variation of the spin-oscillator entanglement entropy for $\omega = 1$ and $\epsilon =5$, as a function of $\delta_{n+1}^2 = (n+1) \gamma^2$. The orange-solid curve represents $S^{\rm I}$ while the black-dotted curve represents $S^{\rm II}$. The blue-dashed vertical line represents the exceptional point.}
    \label{fig_entanglement}
\end{figure}

Resorting now to the Dirac normalization and using the fact that for harmonic-oscillator states, $\langle n|n+1 \rangle = 0$, we can find the reduced density matrix by tracing over the bosonic degree of freedom in the manner
\begin{equation}
{\rm Tr}_{\rm boson} \left[ \frac{|R^{\rm I,II}_{n+1}\rangle\langle R^{\rm I,II}_{n+1}|}{\langle R^{\rm I,II}_{n+1}|R^{\rm I,II}_{n+1}\rangle} \right] = \frac{\left| \frac{1}{2} \right\rangle \left\langle \frac{1}{2} \right| + |\alpha^{\rm I,II}_{n+1}|^2 \left| -\frac{1}{2} \right\rangle \left\langle -\frac{1}{2} \right|}{1 + |\alpha^{\rm I,II}_{n+1}|^2}.
\end{equation}
We thus have the eigenvalues $\lambda^{\rm I,II}_{n+1}$ and $1- \lambda^{\rm I,II}_{n+1}$, with
\begin{equation}
\lambda^{\rm I,II}_{n+1} = \frac{1}{1 + |\alpha^{\rm I,II}_{n+1}|^2}.
\end{equation}
As a result, the entanglement entropy reads
\begin{equation}
S^{\rm I,II} = -  \frac{1}{1 + |\alpha^{\rm I,II}_{n+1}|^2} \ln \left ( \frac{1}{1 + |\alpha^{\rm I,II}_{n+1}|^2} \right) -  \frac{|\alpha^{\rm I,II}_{n+1}|^2}{1 + |\alpha^{\rm I,II}_{n+1}|^2 } \ln \left ( \frac{|\alpha^{\rm I,II}_{n+1}|^2 }{1 + |\alpha^{\rm I,II}_{n+1}|^2} \right),
\end{equation} guaranteed to be real and positive-semidefinite. We can now comment on the behavior of the entanglement entropy pertaining to the unbroken and broken phases:
\begin{enumerate}
\item \textbf{Unbroken phase:} When $\delta_{n+1}^2 < 4$, we have
\begin{equation}
|\alpha^{\rm I,II}_{n+1}|^2 = \frac{(2\mp\sqrt{4 - \delta_{n+1}^2})^2}{\delta_{n+1}^2},
\end{equation} and as a result, $0 \leq S < \ln 2$.  

\item \textbf{Broken phase:} When $\delta_{n+1}^2 > 4$, we have
\begin{equation}
|\alpha^{\rm I,II}_{n+1}|^2 = 1,
\end{equation} and as a result, $S = \ln 2$, i.e., we have maximal entanglement between the spin and the oscillator. 
\end{enumerate}
It can be checked that upon taking $\delta_{n+1}\rightarrow 0$, i.e., in the limit of vanishing spin-oscillator coupling, the entanglement entropy goes to zero consistently. The onset of the quantum phase transition is indicated by the entanglement entropy approaching $\ln 2$ as $\delta_{n+1}^2 \rightarrow (\delta_{n+1}^c)^2 = 4$ at an exceptional point. In other words, when the spin-oscillator entanglement entropy is between 0 and $\ln 2$, the system is in the unbroken phase on the $(n+1)^{\rm th}$ invariant subspace, while the entanglement entropy saturates to $\ln 2$ at the exceptional point and remains saturated throughout the broken phase. The variation of the spin-oscillator entanglement entropy is depicted in Fig. (\ref{fig_entanglement}) which demonstrates our assertions.

\vspace{2mm}

It should be re-emphasized that the entanglement entropy computed here is from a Dirac-normalized density matrix built from the right eigenvectors, meaning that the reduced state obtained by partial tracing is a standard (positive, trace-one) density matrix. If instead one used the left eigenvectors in the Dirac normalization, the reduced spectrum remains unchanged on the two-dimensional invariant subspace. This is because in the basis $\{|n,\frac{1}{2}\rangle, |n+1,-\frac{1}{2}\rangle\}$, the left and right eigenvectors differ only by the phase/sign of the second component (up to normalization), leaving the weights $\{1,|\alpha|^2\}$ unchanged. In general, the entanglement entropies obtained from the Dirac-normalized left and right eigenvectors coincide only when the corresponding reduced density matrices have the same spectrum. This is ensured in the present setting by the fact that the left and right eigenvectors differ only by a relative phase/sign and not by the moduli of their coefficients. For other models, the two reduced spectra need not agree.

\section{Conclusions}
In this work, we have described some interesting properties of a non-Hermitian Jaynes-Cummings model. This system can be realized as a family of infinitely-many closed two-dimensional subspaces, together with a global ground state, the latter being a singlet demonstrating zero entanglement. Each two-dimensional invariant subspace is associated with an exceptional point (for a given $\omega$ and $\epsilon$) where the eigenvalues along with the corresponding eigenvectors coalesce and the system passes from the unbroken phase to the broken phase and vice versa, as illustrated in Fig. (\ref{fig:11}) for the different invariant subspaces. These quantum phase transitions have been interpreted as transitions between the non-dissipative and dissipative regimes, while the exceptional points separating the two phases can be associated with the critical exponent of 1/2, familiar from the theory of exceptional points \cite{Heiss_2012}. On the $(n+1)^{\rm th}$ invariant subspace, we have computed the spin-oscillator entanglement entropy whose distinct profile in the two phases allows one to distinguish between them. While the broken phase is maximally entangled, the entanglement entropy ranges from 0 to $\ln 2$ in the unbroken phase with the lower bound of vanishing entanglement being reached for $\gamma = 0$, in which case the spin and the oscillator do not interact and render the Hamiltonian Hermitian. \\

\noindent
\textbf{Acknowledgements:} G.D. acknowledges UGC-JRF for JRF fellowship. A.G. is grateful to Akash Sinha, Bijan Bagchi, and Miloslav Znojil for related discussions. A.G. has been supported by the Ministry of Education (MoE), Government of India in the form of a Prime Minister’s Research Fellowship (ID: 1200454). A.G. also thanks the Department of Physics, Banaras Hindu University for hospitality.  B.P.M. acknowledges the incentive research grant for faculty under the IoE Scheme (IoE/Incentive/2021-22/32253) of the Banaras Hindu University.

\end{document}